\documentclass[12pt]{iopart}

\usepackage{cite}

\usepackage{graphicx}
\usepackage[dvipdfm,  
            pdfstartview=FitH,
            CJKbookmarks=true,
            bookmarksnumbered=true,
            bookmarksopen=true,
            colorlinks, 
            pdfborder=001,   
            linkcolor=blue,
            anchorcolor=blue,
            citecolor=blue
            ]{hyperref}

\usepackage{fancyhdr}
\usepackage{color}

\begin{document}

\title[Resilience of Epidemics on Networks]{Resilience of Epidemics on Networks}

\author{Lu Dan\textsuperscript{1}, Yang Shunkun\textsuperscript{1}, Zhang Jiaquan\textsuperscript{1}, Wang Huijuan\textsuperscript{2}, Li Daqing\textsuperscript{1,3}}

\address{$^{1}$School of Reliability and Systems Engineering, Beihang University, Beijing, China \\
$^{2}$Intelligent Systems, Delft University of Technology, Delft, Zuid-Holland, Netherlands\\
$^{3}$Science and Technology on Reliability and Environmental Engineering Laboratory, Beijing, China}
\ead{daqingl@buaa.edu.cn}
\vspace{10pt}

\begin{abstract}
Epidemic propagation on complex networks has been widely investigated, mostly with invariant parameters. However, the process of epidemic propagation is not always constant. Epidemics can be affected by various perturbations, and may bounce back to its original state, which is considered resilient. Here, we study the resilience of epidemics on networks, by introducing a different infection rate ${\lambda_{2}}$ during SIS (susceptible-infected-susceptible) epidemic propagation to model perturbations (control state), whereas the infection rate is ${\lambda_{1}}$ in the rest of time. Through simulations and theoretical analysis, we find that even for ${\lambda_{2}<\lambda_{c}}$, epidemics eventually could bounce back if control duration is below a threshold. This critical control time for epidemic resilience, i.e., ${cd_{max}}$ can be predicted by the diameter (${d}$) of the underlying network, with the quantitative relation ${cd_{max}\sim d^{\alpha}}$. Our findings can help to design a better mitigation strategy for epidemics.
\end{abstract}

\pacs{89.75.-k, 87.23.Ge, 64.60.Ht}

\vspace{2pc}
\noindent{\it Keywords}: resilience, epidemics, complex networks

%
%
%
%
%
\maketitle
\makeatletter
\newcommand{\rmnum}[1]{\romannumeral #1}
\newcommand{\Rmnum}[1]{\expandafter\@slowromancap\romannumeral #1@}
\makeatother

{\color{red}\section{Introduction}}

Complex systems~\cite{newman2003structure} in various fields, ranging from natural to engineering systems, such as ecosystems, financial markets, and electric grids, can be viewed as complex networks. Such complex networks~\cite{newman2010networks,cohen2010complex,Caldarelli2007Large} are frequently subject to environmental changes or internal fluctuations. The dynamics on networks~\cite{Barrat2008Dynamical} may possess the capacity to retain the original state essentially after perturbations. Such an adaptive capability is defined as resilience~\cite{Carpenter2001From,Folke2004Regime,Dewenter2014Large,Gao2016Universal,gao2015recent}. For example, it is shown that the food chain~\cite{holling1973resilience} in a biological network can withstand the shocks from a dramatic fall of one species and reorganize into a connected ecological web.

The resilience of epidemics here means that the spreading of epidemics recovers after various perturbations. Most studies on epidemic spreading mainly focus on the phase of the epidemic outbreak on networks~\cite{Bailey1975The,newman2002spread,Pastor2001Epidemic,danon2011networks,Jo2013Enhanced,Li2014Epidemics,Horv2014Spreading,Liu2015Epidemics}. Correspondingly, the parameters characterizing the processes of epidemic transmission on networks are nearly invariant in most mathematical epidemic models~\cite{anderson1992infectious,Hethcote2000The,keeling2005networks}. In some cases~\cite{Meda1969Low,Brockmann2005CHAPTER,Sun2010Influence}, the epidemic spreading may be influenced or reduced by external control of self-repairing mechanism~\cite{Liu2014A,Liu2014Modeling}. However, epidemic may continue to spread by absorbing the perturbations and recover to a stable trajectory, thereby presenting the above-mentioned resilient behaviors.

Resilient behaviors in the context of epidemic propagation have rarely been studied. In this paper, we study the resilience of epidemics on networks based on the classical epidemic model. Three well-known epidemic mathematical models are usually used in the study of epidemic transmission: SI (susceptible-infected) model, SIS (susceptible-infected-susceptible) model and SIR (susceptible-infected-recovered) model~\cite{Bailey1975The,Allen1994Some,Murray1993}. The classical SIS epidemic model is one of the most general way to model the epidemic dynamical behaviors on networks such as Erd\H{o}s-R\'{e}ny\.{i} (ER) networks~\cite{erd6s1960evolution} and scale-free (SF) networks~\cite{barabasi1999emergence}. It is well known that there exists a nonzero epidemic threshold ${\lambda_{c}}$~\cite{Bailey1975The,Marro1999Nonequilibrium} for ER networks in the dynamics of epidemic outbreak. For a given infection rate ${\lambda\geq\lambda_{c}}$, the epidemic will spread out, and the system will reach a stationary state with a finite stable density ${\rho}$ of the infected population. However, if the infection rate ${\lambda}$ is below the epidemic threshold ${\lambda_{c}}$, the epidemic will ultimately die out, with no infected individuals (i.e., ${\rho=0}$). In contrast, it has been demonstrated that the epidemic threshold does not exist for SF networks with ${2<\gamma\leq3}$~\cite{Pastor2001Epidemic}. This has prompted the propagation of viruses in such networks. When the parameter ${\gamma>3}$, the epidemic will spread on these SF networks with a epidemic threshold ${\lambda_{c}}$~\cite{Goltsev2012Localization,Lee2012Epidemic}.

In this paper, we perform studies on the resilience of epidemic transmission with ${\lambda}$ varing over time in the classical SIS epidemic model, to model the perturbations (i.e., "control"). When the epidemic propagation is controlled since a certain time instant (i.e., ${ct}$) for a duration ${cd}$, the infection rate (${\lambda_{2}}$) is smaller than ${\lambda_{1}}$ in the uncontrolled state. Based on different durations of control (i.e., ${cd}$), the transmission level of the epidemic would reflect distinct resilient behaviors after the control. The issue being tackled in this paper is whether the epidemic propagation can present resilient behaviors and the corresponding critical condition.

According to the above-mentioned model, we analyze epidemic spreading on a Facebook network and two types of network models (ER networks and SF networks). The mean-field reaction equation is considered to perform the analytical study of the resilient phenomena, which are in consistent with simulation results. The simulation results have demonstrated that under certain condition, the epidemic can bounce back to the initial steady state in the finite network scale. It is shown on ER networks that even for ${\lambda_{2}<\lambda_{c}}$, the epidemic may eventually bounce back when control time is below a threshold. This is verified by our theoretical analysis of the recovery probability for epidemics, ${P}$, calculated by the equation ${\rho(ct+cd)\rightarrow \frac{1}{N}}$. The critical control time of the resilience (i.e., the critical time for the extinction of epidemics, ${cd_{max}}$), can be predicted by the diameter (${d}$) of networks, with the quantitative relation ${cd_{max}\sim d^{\alpha}}$.

The contents of this paper are arranged as follows. Section~\ref{model} is devoted to introducing the model used to study the resilience of epidemics on networks based on the classic SIS epidemic model. In section~\ref{results}, we perform simulations on different types of networks. Theoretical analysis are also performed on the obtained results. In section~\ref{conclusions}, we draw the conclusions and present discussions.~\\

{\color{red}\section{Model}\label{model}}

In the SIS epidemic model, nodes in the network are divided into two compartments: susceptible individuals (S) and infected individuals (I). Initially, a fraction of nodes, which are randomly selected in the network, are infected. At each time step, each susceptible node is infected by each of its infected neighbors in the network with probability ${\beta}$. Each infected node is cured and becomes susceptible again with probability ${\delta}$ simultaneously. The effective infection rate is defined as ${\lambda=\beta/\delta}$, which is usually constant during epidemic propagation.

In this paper, we study the resilience of epidemics under perturbations, by changing the infection rate ${\lambda}$ to model the perturbations (i.e., "control"). The model applied in networks describing the whole propagation process can be classified into three sub-phases by adding "control", where the starting time and the duration of the "control" are defined as ${ct}$ and ${cd}$, respectively:

Phase 1: The epidemic spreads with an infection rate ${\lambda_{1}=\beta/\delta_{1}}$ lasting ${ct}$ time steps since the beginning ${t=0}$. The probability ${\beta}$ is set to a constant value in the entire process of epidemic transmission.

Phase 2: When the "control" is introduced in the second phase since time ${ct}$, the recovery rate ${\delta_{2}}$ in the controlled network is increased higher than ${\delta_{1}}$, i.e., ${\lambda_{2}}$ is decreased. This process will last for ${cd}$ time steps.

Phase 3: The control is removed at time instant ${ct+cd}$ and the epidemic propagates with the same infection rate, ${\lambda_{1}}$ as in Phase 1.

In addition, the infection densities that we mainly observe are defined as ${\rho_{1}}$, ${\rho_{c}}$ and ${\rho_{r}}$ for each stage. The model can be explained by the following example. In the Internet, some computers may become infected by a certain virus. Accordingly, anti-virus efforts ("control"), i.e., operations including certain immunization process and restoration with anti-virus software~\cite{szor2005art}, will be implemented to stop or lower the epidemic spreading, thus leading to a dramatic decline of the infection rate. When the control is removed due to the limited budgets or adaptation of virus, for different control durations ${cd}$, the epidemic may lose its resilience or resume to spread, therein presenting different resilient behaviors. We study the critical condition by both simulation and theoretical results that the epidemic can continue propagating after the control phase.

We show the results of epidemic resilience on a Facebook network and two types of network models: the ER networks and the scale-free networks. The Facebook network originates from the database on Stanford Large Network Dataset Collection site~\cite{Mcauley2012Learning}. It contains 4039 nodes and 88234 edges, with the average degree of ${\langle k\rangle\approx43.691}$. In addition, the Facebook network has a degree distribution following a power-law distribution with the exponent ${\gamma\approx1.3}$. For an ER network with N nodes, each node pair is independently connected with a probability ${p}$. Then, we create an ER graph with a Poisson degree distribution~\cite{callaway2000network} described as follows:

\begin{eqnarray}\label{eq:ER model}
\ P(k)={N-1\choose k}p^{k}(1-p)^{N-1-k}
\end{eqnarray}

We create scale-free networks via the configuration model~\cite{cohen2010complex}, following a scale-free distribution ${P(k)\sim k^{-\gamma}}$, where ${\gamma}$ is the degree exponent.

In this paper, we first focus on the key quantity about the epidemics, i.e., the infection density ${\rho}$ to study the resilient behaviors of epidemic propagation. Meanwhile, the critical condition for the emergence of the resilient behaviors will be studied. We also explore the probability of recovery for epidemics, ${P}$, as a function of infection rate ${\lambda_{2}}$ in the second phase and control duration ${cd}$. In order to find the critical recovery time of resilience for the epidemic to "bounce back", we calculate the infection density ${\rho_{r}}$ in the third stage, as a function of control duration (${cd}$). In addition, understanding the relation between the resilience of epidemics and network structure is essential to design the resilience strategy. We identify the relation between the critical control time of resilience ${cd_{max}}$ and diameters (${d}$) in ER networks.~\\

\begin{figure}
\par
\centerline{\includegraphics[width=3in]{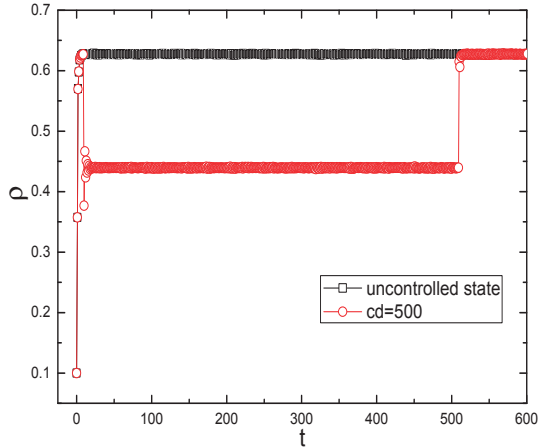}}
\caption{\label{fig:epsart} The infection density ${\rho}$ as a function of time in Facebook network in the uncontrolled state (square) and controlled state with a given control duration (circle). The size of network with the average degree of ${\langle k\rangle\approx43.691}$ is ${N=4039}$. Initially, 10\% of nodes, which are randomly selected in network, are infected. The infection rates are ${\lambda_{1}=\lambda_{3}=0.3}$ for the uncontrolled period and ${\lambda_{2}=0.15}$ for the controlled period, respectively. The starting time of control is ${ct=10}$. The control duration is set to ${cd=500}$. The numerical results are averaged over 100 iterations.}
\par
\label{figure1}
\end{figure}

{\color{red}\section{Results}\label{results}}

To study the resilient behaviors of epidemic propagation, we first perform simulations to obtain the infection density ${\rho}$ as a function of time on different types of networks. To observe the processes of epidemic transmission on a real network, Facebook network is examined with the SIS epidemic model. As shown in figure~\ref{figure1}, the infection density is decreased when the epidemic is controlled with a lower infection rate since time instant ${ct}$. When the control stage is finished and infection rate is recovered, the epidemic can soon restore to the same state before the control.

\begin{figure*}
\centering
\includegraphics[width=5.1in]{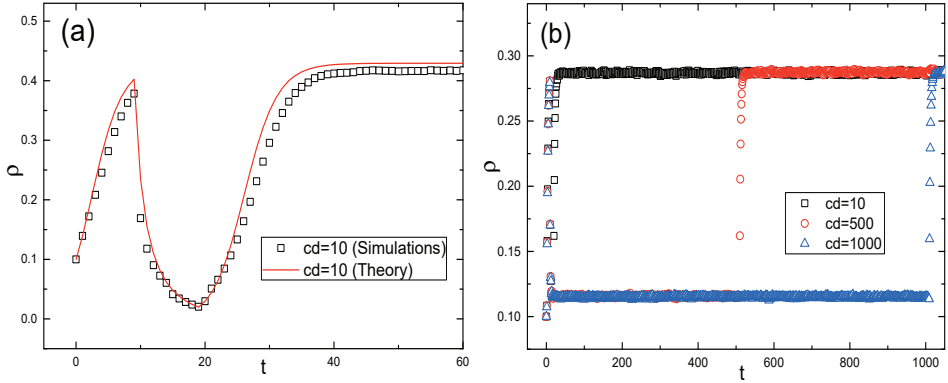}
\caption{\label{fig:epsart} The infection density ${\rho}$ as a function of time in ER and SF networks. (a) Simulation results (splashes) in ER networks with average degree of ${\langle k\rangle=10}$, compared to theoretical values (solid line) as obtained from Eq. (2) and (3). The infection rates are ${\lambda_{1}=\lambda_{3}=0.2}$ for the uncontrolled period and ${\lambda_{2}=0.08<\lambda_{c}=0.1}$ for the controlled period, respectively. The control duration is set to ${cd=10}$. (b) Simulation results (splashes) for SF networks with ${\langle k\rangle\approx 5.4}$ ${(m=2,\gamma=2.5)}$ for different durations of control ${cd=10,500,1000}$. Initially, 10\% of nodes, which are randomly selected in two networks, are infected. The infection rates are ${\lambda_{1}=\lambda_{3}=0.3}$ for the uncontrolled period and ${\lambda_{2}=0.15}$ for the controlled period, respectively. The starting time of control is ${ct=10}$. The size of networks is ${N=10^3}$. The numerical results are averaged over 300 iterations.}
\label{figure2a_2b}
\end{figure*}

Then, we carry out simulations of the epidemic propagation on ER and SF networks. In the case of ER networks, we acquire the results obtained from simulations shown in figure~\ref{figure2a_2b}(a). It shows that the infection density ${\rho}$ is decreased quickly once the "control" is added. When the "control" is removed, the epidemic can also continue to spread and recover to a steady state. It is known that the epidemic threshold ${\lambda_{c}=1/\langle k\rangle}$ through the theoretical calculation~\cite{Pastorsatorras2001Epidemic}, when there is no control in ER networks. Here the infection rate for control phase is smaller than ${\lambda_{c}}$. For SF networks with ${\gamma=2.5}$ illustrated in figure~\ref{figure2a_2b}(b), epidemics can always bounce back to its original state in our simulations at different control durations. This may be due to the fact that the network has no epidemic threshold for ${\gamma<3}$. The findings explicitly demonstrate the existence of the resilient behaviors for the epidemic propagation on SF networks. Meanwhile, the dynamical mean-field rate equation is applied to describe the processes analytically. The first quantity we study here in ER networks is the density of infected nodes ${\rho(t)}$. The governing equation follows:

\begin{eqnarray}\label{eq:equation_first}
\frac{\partial \rho(t)}{\partial t}=(1-(1-\beta)^{\langle k\rangle\rho(t)})(1-\rho(t))-\delta_{1}\rho(t)
\end{eqnarray}

when ${t\leq ct}$, or ${t>ct+cd}$.

\begin{eqnarray}\label{eq:equation_second}
\frac{\partial \rho(t)}{\partial t}=(1-(1-\beta)^{\langle k\rangle\rho(t)})(1-\rho(t))-\delta_{2}\rho(t)
\end{eqnarray}

when ${ct<t\leq ct+cd}$.

It is found that the simulation results on ER networks can be well matched by the theoretical analysis.

In order to find the critical condition for the emergence of the above-mentioned resilient behaviors, we perform theoretical and simulation analysis for the probability of epidemics returning to the steady state in ER networks, ${P}$, as a function of infection rate ${\lambda_{2}}$ in the controlled stage. For theoretical analysis, the probability of recovery for epidemics ${P}$ can be calculated by the mean-field equation written as :~\\

\begin{equation}\label{eq:verify}
\label{cases}
P=\cases{0,& $\rho(ct+cd)\leq\frac{1}{N}$\\
1,& $\rho(ct+cd)>\frac{1}{N}$\\}
\end{equation}

\begin{figure*}
\centering
\includegraphics[width=5.1in]{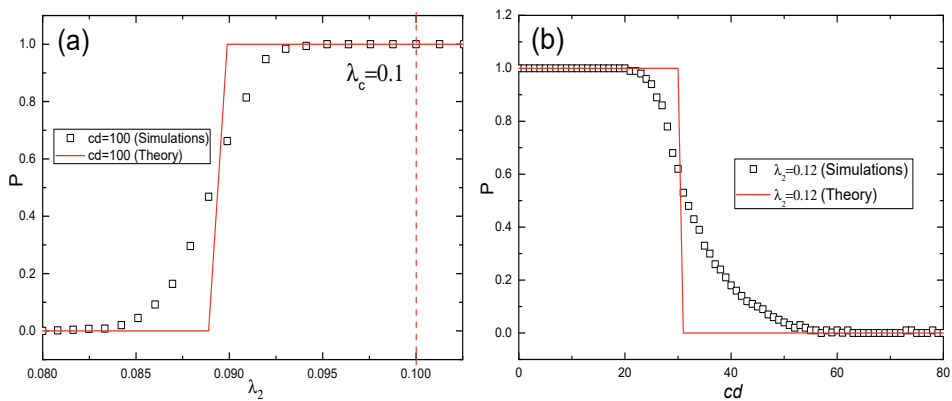}
\caption{\label{fig:epsart} The probability of epidemics returning to the steady state ${P}$ as a function of the infection rate ${\lambda_{2}=\beta/\delta_{2}}$ in control stage shown in (a) and as a function of control duration (${cd}$) shown in (b). Simulation results (splashes) for ER networks with size of ${N=10^4}$, compared to theory (solid line). (a) ER networks with ${\langle k\rangle=10}$. The control duration is set to ${cd=100}$. The infection rate is ${\lambda_{1}=\lambda_{3}=0.2}$ in uncontrolled stage. The dash line represents the value of epidemic threshold, i.e., ${\lambda_{c}=1/10}$. (b) ER networks with ${\langle k\rangle=6}$. The infection rates are ${\lambda_{1}=\lambda_{3}=0.3}$ for the uncontrolled period and ${\lambda_{2}=0.12<\lambda_{c}=0.167}$ for the controlled period, respectively. Initially, 10\% of nodes, which are randomly selected, are infected. The starting time of the control is ${ct=10}$.}
\label{figure3a_3b}
\end{figure*}

In figure~\ref{figure3a_3b}(a), the probability of epidemic restoration, ${P}$, grows to 1.0 eventually with increasing ${\lambda_{2}}$ for a given ${cd}$. There exists a critical infection rate ${\lambda^{c}_{2}}$ in control stage for epidemic resilience. Meanwhile, even for ${\lambda_{2}<\lambda_{c}}$, where ${\lambda_{c}}$ is the epidemic threshold (${\lambda_{c}=1/\langle k\rangle}$) shown by dash line, the epidemic may still bounce back. When the equation satisfies ${\rho(ct+cd)\leq\frac{1}{N}}$ in theory, the epidemic is considered to lose the resilience. The theoretical results, based on the above-mentioned calculation method, can predict the critical ${\lambda^{c}_{2}}$, verified by the simulation results. It illustrates that the epidemics will bounce back with ${\lambda_{2}\geq\lambda^{c}_{2}}$ (${\lambda_{2}<\lambda_{c}}$) for a given control duration ${cd}$. This threshold for epidemic resilience depends on the combined effect of epidemic spread and control processes.

To explore the effect of the control duration ${cd}$ on the probability for epidemics to return to the steady state, we perform the theoretical and simulation analysis for ${P}$ as a function of ${cd}$ in ER networks. As shown in figure~\ref{figure3a_3b}(b), the probability ${P}$ is decreased with the increasing ${cd}$ for a given infection rate ${\lambda_{2}}$ (${<\lambda_{c}=0.167}$). When the control duration (${cd}$) is small, the probability maintains 1.0. As ${cd}$ increases, the recovery probability is decreased and finally reaches to zero. It is shown that there exists a threshold (i.e., ${cd_{max}}$) that makes the epidemic lose the resilience completely. And ${cd_{max}}$ can also be regarded as the critical recovery time of resilience. The theoretical results (solid line) are obtained based on the equation (\ref{eq:verify}). Our theoretical analysis can also predict the critical control duration, ${cd_{max}}$, above which the epidemic will lose its resilience.

\begin{figure*}
\centering
\includegraphics[width=5.1in]{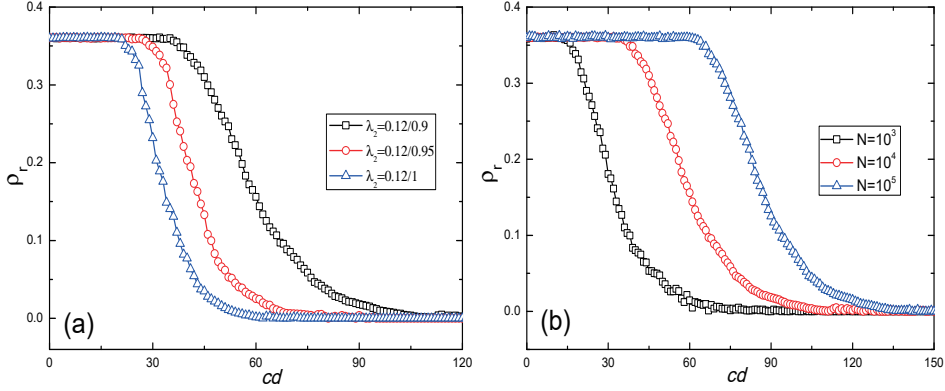}
\caption{\label{fig:wide}The infection density after control stage, ${\rho_{r}}$, obtained in the steady state, as a function of control duration (${cd}$). (a) ER networks with ${\langle k\rangle=6}$, ${N=10^{4}}$. The infection rate in a controlled state is ${\lambda_{2}=0.12/0.9,0.12/0.95,0.12/0.1}$. (b) ER networks with ${\langle k\rangle=6}$ by setting ${N=10^{3},10^{4},10^{5}}$, given ${\lambda_{2}=0.12/0.9}$. Initially, 10\% of nodes, which are randomly selected, are infected. The starting time of the control is ${ct=10}$. The infection rate is ${\lambda_{1}=\lambda_{3}=0.3}$ in uncontrolled stage. The results above have been averaged over 300 realizations.}
\label{figure4a_4b}
\end{figure*}

As the resilience for the epidemic propagation are affected by the infection rate ${\lambda_{2}}$ in the second stage of transmission and the control duration (${cd}$), we study the resilience output with the steady-state density in the third stage, ${\rho_{r}}$, as a function of ${cd}$. As shown in figure~\ref{figure4a_4b}(a), for a fixed ${\lambda_{2}}$, the infection density ${\rho_{r}}$ is decreased as ${cd}$ is increased for an ER network. When ${cd}$ is increased to a certain value, ${\rho_{r}}$ is almost reduced to zero. This confirms our finding for ${cd_{max}}$ in figure~\ref{figure3a_3b}. It can also be seen that ${\rho_{r}}$ falls to zero earlier with a smaller infection rate ${\lambda_{2}}$ than that with a large ${\lambda_{2}}$. To study the effect of network size on the resilience of epidemics, we perform simulations to obtain ${\rho_{r}}$ as a function of ${cd}$ in ER networks with different sizes. As shown in figure~\ref{figure4a_4b}(b), in the case of ER networks, ${\rho_{r}}$ is reduced to zero gradually with increasing ${cd}$ time steps for various ${N}$. Infection density after control stage, ${\rho_{r}}$, is decreased more quickly with a relatively small ${N}$. It is considered that ${cd_{max}}$ is larger for a large ${N}$.

\begin{figure*}
\centering
\includegraphics[width=3in]{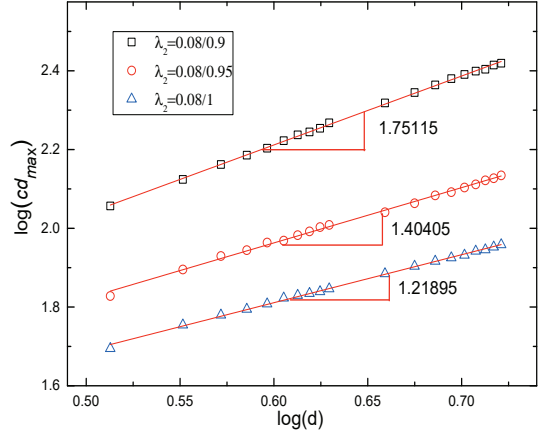}
\caption{\label{fig:epsart} The threshold ${cd_{max}}$ as a function of diameters (${d}$) for ER networks with ${\langle k\rangle=10}$. The infection rate is given as ${\lambda_{2}=0.08/0.9,0.08/0.95,0.08/1}$ in control stage. Initially, 10\% of nodes, which are randomly selected, are infected. The starting time of the control is ${ct=10}$. The infection rate is ${\lambda_{1}=\lambda_{3}=0.2}$ in uncontrolled stage. The results are averaged over 100 realizations.}
\label{figure5}
\end{figure*}

It is meaningful to understand the relation between epidemic resilience and network structure, which can help to design the resilience strategy. Therefore, we perform simulations for ${cd_{max}}$ as a function of diameters (${d}$) in ER networks. In figure~\ref{figure5}, for a given ${\lambda_{2}}$, the ${cd_{max}}$ is found to scale with network diameters in ER networks, ${cd_{max}\sim d^{\alpha}}$, where ${\alpha}$ is increasing with increasing ${\lambda_{2}}$. This is because that a larger ${\lambda_{2}}$ requires a longer duration of control for the recovery of resilience.

{\color{red}\section{Conclusions}\label{conclusions}}

By adding a "control" stage in the original SIS model, we model the resilience of epidemic propagation under perturbations. When the network is in the controlled state, the epidemic transmits with a smaller infection rate ${\lambda_{2}}$ than that in the uncontrolled state. When the control is removed, the epidemic may restore to a steady state exhibiting resilient behaviors.

Based on the above-mentioned model, we performed numerical simulations on a Facebook network and two types of network models (ER and SF networks). The simulation results indicate that under certain condition, the epidemic can restore to the original steady state in the finite network size. The mean-field reaction equation is applied to perform the analytical study of the resilient phenomena, which are in consistent with simulation results. Through the simulations and theoretical analysis on ER networks, it is shown that even for ${\lambda_{2}<\lambda_{c}}$, the epidemic may eventually bounce back when the control duration, ${cd}$, is smaller than a threshold ${cd_{max}}$. It can be verified by the theoretical results of the recovery probability for epidemics, ${P}$, computed by the equation, ${\rho(ct+cd)\rightarrow\frac{1}{N}}$. The critical value ${cd_{max}}$ is strongly related to the network structure, where ${cd_{max}}$ can be predicted by the diameter ${d}$ of networks with the quantitative relation, written as ${cd_{max}\sim d^{\alpha}}$. Note that the maximum time of control for epidemics extinction is increasing with system size and will diverge for an infinite system. The resilience is an intrinsic property for epidemics to adapt to the external perturbations and the changes of internal conditions. It can also issue signals for systems to mitigate the epidemic rapidly and accurately when they have been infected. The discovery of ${cd_{max}}$ may provide advanced indication for the resilience of the epidemic, which can help to design protection strategy keeping systems from a secondary epidemic outbreaks.~\\

{\color{red}

\end{document}